\documentclass[epj]{svjour}
\usepackage{graphics}
\begin{document}
\title{Nonstationary stochastic resonance viewed through the lens
of information theory}
%\subtitle{}
\author{Igor Goychuk and  Peter H\"anggi}
%\thanks{\emph{Present address:} }%
%}                     % Do not remove
%
\offprints{I. Goychuk}          % Insert a name or remove this line
\institute{Institute of Physics, University of Augsburg,
   Universit\"atsstr. 1, D-86159, Augsburg, Germany
          %\email{goychuk@physik.uni-augsburg.de}
      }
\date{Received: date / Revised version: date}
\abstract{ In biological systems, information is frequently
transferred with Poisson like spike processes (shot noise) modulated
in time by information-carrying signals. How then to quantify
information transfer  for the output for such nonstationary input
signals of finite duration? Is there some minimal length of the
input signal duration versus its strength? Can such signals be
better detected when immersed in noise stemming from the
surroundings by increasing the stochastic intensity? These are some
basic questions which we attempt to address within an analytical
theory based on the Kullback-Leibler information concept applied to
random processes. }

 % end of PACS codes
%} %end of abstract
%
\maketitle
\section{Introduction}
\label{intro}

Stochastic resonance (SR) \cite{Benzi} grew into extensive  research
domain on the border between many scientific disciplines, ranging
from geophysics and climate dynamics, numerous physical, biophysical
and engineering applications \cite{Review98,Review99,Hanggi},
including  quantum SR in the deep quantum cold \cite{QSR}.  Nevertheless,
more complex physical SR applications and timely SR applications to
biological and climate complexity \cite{rahmstorf} as well as more
insightful reasoning are still in the limelight. The original
statement of the problem, i.e. a paradoxical amplification of the
signal  in a noisy background due to intrinsic \cite{intrinsicSR},
or added, external  noise \cite{Review98,Review99,Hanggi}, has been
contrasted with a synchronization framework
\cite{Review98,Review99,Hanggi,Freund}. {\it Postfactum} we can
reformulate the original problem by asking the question of whether a
stochastic bistable clock can resonate with an externally applied
periodic driving via increasing the randomness of  the underlying
bistable clock dynamics.

The notion of a ``Stochastic clock'' \cite{Kiss} stems  
conceptually from the
theory of continuous time random walk processes [see e.g. in Ref.
\cite{Hughes}, p. 245]. It is characterized by a distribution of the
sum of stochastic periods. The distribution of one period duration
 is a convolution of the residence times in the two clock states.
 Two subsequent transitions perform a cycle with a random duration.
 It is important to note   that if the mean duration of a cycle
 $\langle \tau_{cycle}\rangle$ exists, the distribution of $n$
 cycles duration yields a sharp function, centered at
 $n\langle \tau_{cycle}\rangle$ in the limit $n\to\infty$.
 Intrinsic noise changes $\langle \tau_{cycle}\rangle$ and
 in some situations, -- e.g. for symmetric Markovian clock with an exponential
 distribution of the residence times and an exponential
 dependence of the mean cycle durations on the noise intensity --, the
 stochastic clock can resonate with a weak periodic driving of period
 ${\cal T}_0$, when $\langle \tau_{cycle}\rangle={\cal T}_0$. This
 is the benchmark of the stochastic resonance phenomenon.
 Then, the periodic signal is best
detectable
 in the spectral power spectrum of the clock's bistable
 fluctuations  and the stochastic transitions become
 more correlated with the periodic  time course of the signal.

 Whether the information transfer will be
 optimized at this resonance condition depends on how is information
 encoded. If a direct encoding is used, i.e. {\it locally}
 in the time domain, then
 the answer is ``yes''. However, if information is encoded in the frequency domain (like
used
 in radio devices), then for the discussed bistable clock the answer
 is typically ``no'', at least for weak signals.
 The spectral signal-to-noise ratio (SNR) characterizes the Shannon
 information transfer for weak signals \cite{Shannon} (more precisely,
 the information capacity of an information channel
 which is the maximal rate of the Shannon's
 mutual information between the input and output signals for the fixed
 total power of the input signal). Stochastic resonance in the
 spectral SNR for the stochastic bistable system does  not necessarily reflect a synchronization phenomenon
 \cite{Review98,Review99,Hanggi,Freund}. More specifically, SNR does
 not directly reflect the matching between the stochastic time scale of the bistable
 clock dynamics and the time-scale of the deterministic, coherent signal dynamics.
 In contrast to SNR, the measure of spectral amplification
 \cite{junga,jungb}, however,
 explicitly involves a dependence on the driving frequency.
 We also remark while within linear response (i.e. weak signals)
 the SR gain determined by the ``SNR output/ SNR input'' cannot
 exceed unity \cite{Casado03} this  is no longer the case for the
 nonlinear SR response \cite{Casado03,saga}.

Next we  are dealing with SR in a
 wider sense, i.e. we shall study  SR in a form which can broadly be characterized as a {\it relative}
 amplification of the information transfer
 through a noisy system. SNR for a weak sinusoidal signal predicts the information transfer
for stationary, weak broadband stochastic signals
\cite{Shannon,Spikes,Bialek}. As just noted, this remarkable analogy
fails, however, for strong signals beyond the linear response
approximation \cite{Casado03,saga}. Nevertheless,  it must be
remarked that it is simply not feasible to extract more information
(in the Shannon sense) from the output signal than  was originally
encoded in the input signal; this agrees with common sense and is
corroborated with  the  information processing 
inequality \cite{Shannon}.

\begin{figure}
\resizebox{0.75\columnwidth}{!}{%
  \includegraphics{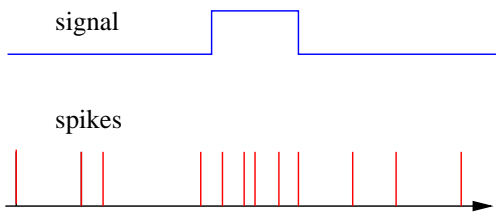}
}
%\vspace{5cm}       % Give the correct figure height in cm
\caption{(Color online) A step-like signal modulates an
information-carrying spiking process, e.g. output of a sensory
neuron \cite{Koch}.}
\label{Fig1}       % Give a unique label
\end{figure}

 In the case of so-termed {\it aperiodic SR}, i.e. SR fed by  stationary
 stochastic input signals (modeled, e.g., by Gaussian processes)
 the rate of mutual information can be used as a suitable
 quantifier \cite{Collins}. How to proceed,
 however, if the signal is not stationary as it is
 intrinsically the case for fixed, deterministic or stochastic  non-stationary inputs of  finite duration?
 This latter situation is typical, e.g., for  biological systems \cite{Review98,Review99,Hanggi},
 cf. a typical situation depicted with
  Fig. \ref{Fig1}.
Typical spectral measures as indicated above are then of limited
use, or at best of approximate use only. The mutual information
concept also cannot  be applied whenever the input signal is
strictly deterministic. In such a nonstationary situation one can
characterize the information
 transfer by the change of the entropy between the process in absence of input and the output process
 when the input signal is applied. Then the difference of entropies can be regarded as
 the information gained from the input signal \cite{Shannon}.
 The Kullback-Leibler {\it relative} entropy 
 \cite{Kullback,MacKayBook,Neiman},
 termed also the information gain is a  suitable measure to characterize 
 the corresponding entropy difference
 because it does not suffer from the subjective dependence
 on the discrete time-step $\Delta \tau$ used in approximating   continuous  time
 random processes \cite{GH00,Goychuk01}.  
 This presents an advantageous fact
 when  contrasted with using   direct entropy differences \cite{Spikes}.
 The Kullback-Leibler entropy is just an analogue of the
 Boltzmann $H-$function for kinetic equations \cite{Kampen}.
 In the present context, it is
 applied not just to a single
 time probability density, but rather
 to the whole probability {\it functional} that determines the stochastic process under consideration.
 Like the $H-$function it characterizes the entropy difference from
 equilibrium in a well-defined manner, even if the
 equilibrium value of informational entropy itself is not
 precisely defined for continuous
 distributions. This is always the case  if
  some fundamental ``quantum-scale'' (like the Planck constant for an
  elementary ``area'' in the phase space of a physical system, in the
  case of physical entropy) is absent.
 Moreover, the use of information gain if averaged over all possible
 realizations of a random input signal provides an upper bound for the mutual
 information transferred \cite{GH00,Goychuk01}.
 These facts  predestine indeed
 the information gain as an adequate measure to characterize  nonstationary SR.

 With this present work, we investigate such nonstationary SR for an archetype
 setup of  SR  \cite{Review98,Goychuk01,Model,Bezrukov},
 by considering the renewal point processes for
  signal-modulated spike occurrences. This sort of modeling is  relevant
 to signaling occurring in biological systems  \cite{Spikes,Koch,Julicher,Skupin}.

 \section{The model}

\begin{figure}
\resizebox{0.75\columnwidth}{!}{%
  \includegraphics{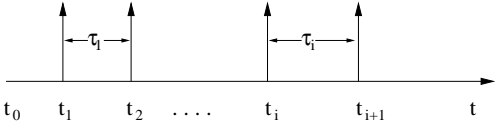}
}
%\vspace{5cm}       % Give the correct figure height in cm
\caption{Driven renewal  process: spikes occur at random times $t_i$,
the interspike time-intervals $\tau_i=t_{i+1}-t_i$ are assumed
to be uncorrelated and described by a non-homogeneous 
ISI density $\psi(\tau|t)$.
}
\label{Fig2}       % Give a unique label
\end{figure}

We consider a renewal point process $\xi(t)$ defined by the  spikes
\begin{eqnarray}
 \xi(t)=\sum_i \xi\delta(t-t_i)
\end{eqnarray}
occurring at random times $t_i$, see in  Fig. \ref{Fig2}. One may
safely assume that the spike width is negligible and its form is
fixed by some total intensity (the time integral of the spike-form)
$\xi$, i.e. the information is transferred with the spike
occurrences. Put differently, rather than a specific shape of the
spike it is the timing dynamics and not only the averaged number
that is important in transferring information. The
interspike intervals (ISI) $\tau_i=t_{i+1}-t_{i}$ are assumed to be
uncorrelated (renewal assumption) and the whole process, generally a
non-Markovian process,  is completely characterized by the ISI
distribution $\psi(t+\tau,t):=\psi(\tau|t)$, or equivalently by the
corresponding survival probability $\Phi(t+\tau,t):=\Phi(\tau|t)$,
$\psi(\tau|t)=-d \Phi(\tau|t)/d\tau$ \cite{GH04,remark}. The process
is {\it non-homogeneous} in time what is reflected by its explicit
dependence of the above quantities on the current time $t$ via an
input  signal $V_s(t)$. For a time-homogeneous process we have in
contrast $\psi(\tau|t)=\psi(\tau)$. The simplest example is given by
the Poisson process with the time-dependent rate $r(t)$
\cite{Kampen,Papoulis}, where
\begin{eqnarray}
 \Phi(\tau|t)=\exp\left (-\int_{t}^{t+\tau}r(t')d t'\right )\;.
\end{eqnarray}
A popular SR model is \cite{Review98,Bezrukov,Model}:
\begin{eqnarray}\label{model1}
r(t)=r_0(U_0)\exp(qV_s(t)/D)\;,
\end{eqnarray}
where
\begin{eqnarray} \label{model2}
r_0(U_0)=k_0\exp(-U_0/D)
\end{eqnarray}
is the rate in the absence of signal. One further assumes that there
is a threshold $U_0$ which upon crossing induces a spike. Typical
realizations are a conventional  threshold detector \cite{Gingl}, or
the dynamics of the leaky integrate-and-fire model of neuron firing
\cite{Koch} driven, e.g., by synaptic noise. Some intrinsic noise of
the intensity $D$ produces spontaneous firing and the signal
modulates the threshold height. A similar model applies also to the
current spikes produced in a membrane by spontaneous electroporation
facilitated by some ion channel forming peptides \cite{Bezrukov}. In
the latter case the signal $V_s(t)$ is the voltage modulation, the
quantity $q$ is an effective gating charge, $U_0$ is the energy
barrier to the channel formation, and $D=k_BT$. A similar model
(modified for the refractory times) provides also a crude
approximation to the activity of cortical and sensory neurons
\cite{Koch}, spontaneous spiking of the ion channel clusters
\cite{ChowWhite}, and spontaneous calcium release spiking in living
cells \cite{Skupin}.

\section{Theory}

We start out from the trajectory description of $\xi(t)$ considering
a finite time interval $[t_0,t)$. The elements of the probability
space are the trajectories with spikes occurring
%$n=0,1,2,...$ spikes occurring
at some random times $t_1,t_2,...,t_n$. The probability to have
spike occurring during the prescribed  interval (starting out from a
no spike event at time $t_0$ ) is $P_{0}(t,t_0)=\Phi(t,t_0)$. The
probability density of trajectories with one spike occurring at
$t_1$, $t_0<t_1<t$ is
\[Q_{1}(t,t_1,t_0)=\Phi(t,t_1)\psi(t_1,t_0),\]
yielding for the corresponding probability
\[P_{1}(t,t_0)=\int_{t_0}^t Q_{1}(t,t_1,t_0)dt_1\;.\]
Furthermore, the probability density of trajectories with two events
at $t_1$ and $t_2$, $t_0<t_1<t_2<t$, reads
\[Q_{2}(t,t_2,t_1,t_0)=\Phi(t,t_2)\psi(t_2,t_1)
 \psi(t_1,t_0), \]
and the probability to have two events within $[t_0,t)$ becomes
\[P_{2}(t,t_0)=
\int_{t_0}^t dt_2\int_{t_0}^{t_2} dt_1 Q_{2}(t,t_2,t_1,t_0)\;.\]
Other probability densities and probabilities are constructed akin,
using the semi-Markov, renewal character of the underlying process.
The normalization condition \\
$\sum_{n=0}^{\infty}P_n(t,t_0)=1$ can be
readily verified; it is done by  showing that the derivative of the
l.h.s. with respect to $t$ is zero upon using $P_0(t_0,t_0)=1$,
$P_{n\neq 0}(t_0,t_0)=0$. We  thus obtain a complete description of
the considered
time-inhomogeneous, nonstationary  process with the probability density functional \\
\[P[\xi(t)]=
\left [P_0(t,t_0), Q_1(t,t_1,t_0), ..., Q_n(t,t_n,...,t_1,t_0), ...
\right ].\] For this rate-modulated Poisson process the densities
read
\begin{eqnarray}\label{distribution}
Q_n(t,t_n,...,t_1,t_0)=
\exp\left(-\int_{t_0}^{t}r(t')dt'\right)\prod_{i=1}^n r(t_i),
\end{eqnarray}
and the number of spikes exhibits a Poisson distribution
\begin{eqnarray}
P_n(t,t_0)=\frac{\langle n(t,t_0)\rangle^n}{n!}
\exp\left [ -\langle n(t,t_0)\rangle \right ]\;
\end{eqnarray}
with mean $\langle n(t,t_0)\rangle =\int_{t_0}^{t}r(t')dt'$.

\subsection{$\tau-$Entropy}

The definition of the entropy of any continuous variable which has a
physical dimension requires to introduce some arbitrary  bin $\Delta
\tau$ (a measurement unit). The entropy  of $\xi(t)$ can be defined
(in natural units, nats) as a functional integral \cite{Gaspard}
\begin{eqnarray}\label{def}
S_{\Delta \tau}(t,t_0)& = & -P_0(t,t_0)\ln P_0(t,t_0) \nonumber \\
& - &\sum_{n=1}^{\infty}
\int_{t_0}^{t}dt_n\int_{t_0}^{t_n} dt_{n-1}...\int_{t_0}^{t_2}dt_1  \\
&& Q_n(t,t_n,...,t_0) \ln [Q_n(t,t_n,...,t_0)(\Delta \tau)^n] \;.\nonumber
\end{eqnarray}
Upon differentiating Eq. (\ref{def}) with respect to time $t$ we
arrive after some algebra at the following expression for the rate
of entropy production
\begin{eqnarray}\label{exact}
&&\frac{d}{dt} S_{\Delta \tau}(t,t_0) =
F(t,t_0)+\int_{t_0}^{t}F(t,t_1)\Psi(t_1,t_0)dt_1\; \\
&& =F(t,t_0)+\sum_{n=1}^{\infty}
\int_{t_0}^{t}dt_n\int_{t_0}^{t_n} dt_{n-1}...\int_{t_0}^{t_2}dt_1
\frac{F(t,t_n)}{\Phi(t,t_n)} \nonumber \\ && \times Q_n(t,t_n,...,t_0) \;. \nonumber
\end{eqnarray}
In Eq. (\ref{exact}),
\begin{eqnarray}\label{aux1}
F(t,t_1):=\dot \Phi(t,t_1)\ln\left(\frac{-\Delta \tau
\dot \Phi(t,t_1)}{
e\Phi(t,t_1)}\right)
\end{eqnarray}
where $\dot \Phi(t,t_1)\equiv d\Phi(t,t_1)/dt<0$ and
\begin{eqnarray}
\Psi(t,t_0)  = \sum_{n=1}^{\infty}
\int_{t_0}^{t}dt_n\int_{t_0}^{t_n} dt_{n-1}...\int_{t_0}^{t_2}dt_1
\nonumber \\
\times
\prod_{i=1}^{n}\psi(t_i,t_{i-1})\;.
\end{eqnarray}
Using  that
\[ \Phi(0|t)=1, \;\;\lim_{\tau\to\infty}
\Phi(\tau|t)=0,\] one can show that $S_{\Delta\tau}(\psi|t)=
\int_0^{\infty}F(t+\tau,t)d\tau$ has the meaning of the entropy of
the time-inhomogeneous ISI, i.e.
\begin{eqnarray}\label{s-tau}
S_{\Delta\tau}(\psi|t)=-
\int_0^{\infty}\psi(\tau|t)\ln[\psi(\tau|t)\Delta
\tau]d\tau\;.
\end{eqnarray}

For the Poisson model, $\Psi(t,t_0)=r(t)$
and Eq. (\ref{exact}) simplifies
to
\begin{eqnarray}\label{rate1}
\frac{d}{dt}S_{\Delta \tau}(t,t_0)=r(t) \ln \left(\frac{e}{r(t)\Delta \tau
} \right)=r(t)\overline{S}_{\Delta \tau}(\psi|t),
\end{eqnarray}
where $\overline{S}_{\Delta \tau}(\psi|t)$ is the entropy of the ISI
distribution which is calculated with the {\it frozen} rate $r(t)$,
i.e. with $\psi(\tau)=r(t)\exp(-r(t)\tau)$ instead of $\psi(\tau|t)$
in Eq. (\ref{s-tau}). This result has a simple interpretation:
namely that the {\it rate of entropy production = spiking rate
$\times$ entropy of ISI distribution for instant rate $r(t)$}. For
the ``background'' process, i.e. the resulting process  with  no
signal applied,  the rate is $r_0$ and the entropy of the spike
train of duration ${\cal T}$ is given by the well-known MacKay and
McCulloch result \cite{Spikes,MacKay}
\begin{eqnarray}\label{S0}
S_{0}= N({\cal T})\ln\left (\frac{e}{r_0\Delta\tau}\right),
\end{eqnarray}
where $N({\cal T})=r_0{\cal T}$ is the averaged number of spikes. A
popular definition of the information $I_{\Delta\tau}({\cal T})$
transferred with spikes \cite{Spikes,Bialek} amounts to take the
difference
\begin{eqnarray}
&&I_{\Delta\tau}({\cal T})=S_0-S_{\Delta\tau}(t_0+{\cal T},t_0)\\
&&=\int_{t_0}^{t_0+{\cal T}} \left [
r_0\ln\left (\frac{e}{r_0\Delta\tau}\right)-
r(t)\ln\left (\frac{e}{r(t)\Delta\tau}\right)\right ]dt \nonumber.
\end{eqnarray}
One can see  that the dependence on the finite  time bin $\Delta
\tau$ does  generally not cancel \cite{Goychuk01,GH04}, unlike in
the case  of an $n$-dimensional probability distribution. The reason
is that the probability $P_n$ to have $n$ spikes is changed, i.e.
probability is redistributed between different $n$-dimensional
``slices'' of the hyper-dimensional probability space.
%This arbitrary $\Delta\tau-$dependence can yet be interpreted
%\cite{Spikes}.
Troublesome is further the finding that the above difference can
readily  become negative (i.e. for  $r(t)>r_0$ and  for a
sufficiently small $\Delta\tau$). This should then be interpreted as
a loss of information. Of course, a proper definition should always
yield a positive information, because the spikes become more ordered
due to the application of the input signal. The interpretation
problem is an artefact of this  $\Delta \tau$ dependence. This is
precisely  why we  prefer to define the entropy difference for
stochastic processes via the Kullback-Leibler relative entropy, see
below. Before we proceed with a suitable definition along our posed
objective we derive next a generalization of the MacKay and
McCulloch result for different, nondriven (i.e. $V_{s}(t) = 0$)
point processes.

\subsection{Rate of entropy production for stationary  renewal point processes}

Using the exact result (\ref{exact}) one can find also the
asymptotic rate of entropy production for time-homogeneous processes
in the limit $t\to\infty$ for any $\psi(\tau)$ with a {\it finite}
mean ISI $\langle \tau\rangle=\int_0^{\infty}\tau\psi(\tau)d\tau$.
In this case $F(t+\tau,t)=F(\tau)$, $\Psi(t+\tau,t)=\Psi(\tau)$, and
for the Laplace-transformed rate of the entropy production $\dot
S_{\Delta\tau}(t,0)$ we obtain
\begin{eqnarray}
 \tilde R(s)=\frac{\tilde F(s)}{1-\tilde\psi(s)}\;.
\end{eqnarray}
The asymptotic rate of entropy production follows as
\begin{eqnarray}
\lim_{t\to\infty} \frac{d}{dt}S_{\Delta \tau}(t,0) =
\lim_{s\to 0}[s\tilde R(s)]=
\frac{1}{\langle
\tau\rangle} S_{\Delta \tau}(\psi)\;,
\end{eqnarray}
which is a natural generalization of the relation (\ref{rate1}). The
role of the mean spiking rate is taken on by $1/\langle\tau\rangle$.
The result in (\ref{S0}) is thus generalized to read
\begin{eqnarray}\label{S0new}
S_{0}= N({\cal T})S_{\Delta \tau}(\psi),
\end{eqnarray}
where $N({\cal T})={\cal T}/\langle \tau\rangle$. This latter result
is applicable also in the case of fractal-rate renewal processes
\cite{Lowen}, where the mean rate does not exist, i.e. where
$\tilde\psi(s)\approx 1-(s\tau^*)^{\alpha}$, for $s\to 0$;
$0<\alpha<1$ and $\tau^*$ is some scaling time parameter
\cite{Hughes}. The only difference is that the number of spikes
within a long time interval ${\cal T}$ scales sub-linearly with its
length, i.e. $N({\cal T})\propto ({\cal T}/\tau^*)^{\alpha}$.
Generally, for the considered renewal processes we find that {\it
entropy of spike train = number of spikes $\times$ entropy of ISI}.
One can also infer that the Poisson process is the maximum entropy
point process, for  fixed $\langle\tau\rangle$ and $\Delta\tau$.
This is so, because the exponential ISI distribution displays the
maximum entropy distribution from all one-sided distributions under
such constraints.

% not   the focus of our work ...too long section already ,.....
%This fact can rationalize the use of the Poisson like processes by
%Nature for the information transduction: the background
%information-carrying process should be either a maximally ordered,
%zero-entropy process, e.g. a periodic wave, or a maximally
%disordered process. Both ways, the {\it entropy difference} due to a
%signal-modulation, i.e. the transferred {\it information} about
%signal, can be maximized.

\subsection{Kullback-Leibler relative entropy}

\begin{figure}
\begin{center}
\resizebox{0.63\columnwidth}{!}{%
  \includegraphics{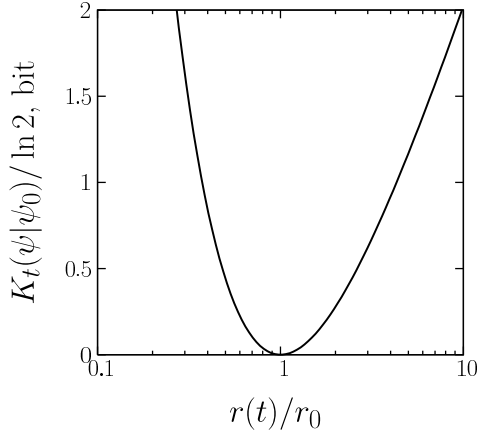}
}
\end{center}
\vspace{1cm}
%\vspace{5cm}       % Give the correct figure height in cm
\caption{Kullback-Leibler entropy per one spike as a function
of the frozen rate $r(t)$.}
\label{Fig3}       % Give a unique label
\end{figure}

As discussed above, $\tau$-entropy of the spike train is 
not exactly defined, being 
dependent on $\Delta \tau$.  However, its deviation from
equilibrium can be defined unambiguously via the relative entropy,
given by  the  functional integral:

 \begin{eqnarray} \label{kulldef}
&&K_{[t_0,t]}[\xi(t)|\xi_0(t)] \nonumber \\
= &&
\int D[\xi(t)] P[\xi(t)|V_s(t)]
\ln \left(
\frac{P[\xi(t)|V_s(t)]}{P[\xi_0(t)]}
\right)  \nonumber \\
  = && P_0(t,t_0|V_s)\ln\left(
 \frac{ P_0(t,t_0|V_s)}{P_0^{(0)}(t,t_0)}\right) \nonumber \\
  +&&   \sum_{n=1}^{\infty}%\frac{1}{n!}
\int_{t_0}^{t}dt_n\int_{t_0}^{t_n} dt_{n-1}...\int_{t_0}^{t_2}dt_1
Q_n(t,t_n,...,t_0|V_s) \nonumber \\
  \times && \ln \left(\frac{Q_n(t,t_n,...,t_0|V_s)}
 {Q_n^{(0)}(t,t_n,...,t_0)}\right) \geq 0  \;.
\end{eqnarray}
Here, the super-index $(0)$ refers to the background process $\xi_0(t)$ in
the absence of signal $V_s(t)$.
Relative entropy is always non-negative. It is zero iff the both
distribution functionals $P[\xi_0(t)]$ and $P[\xi(t)|V_s(t)]$ coincide \cite{Kullback},
i.e. in the absence of signal. The rate of the relative entropy production
for the rate-modulated Poisson process can be easily calculated with Eq.
(\ref{distribution}) in (\ref{kulldef}) \cite{Goychuk01}. It reads,
\begin{eqnarray}\label{result}
\frac{d}{dt}K_{[t_0,t]}[\xi(t)|\xi_0(t)]=r(t)  K_t(\psi|\psi_0)\geq 0,
\end{eqnarray}
where
\begin{eqnarray}\label{rel}
K_t(\psi|\psi_0)&=&\int_{0}^{\infty}d\tau\psi(\tau)
\ln\left(\frac{\psi(\tau)}{\psi^{(0)}(\tau)} \right) \nonumber \\
&=&\frac{r_0}{r(t)}-1+\ln\left( \frac{r(t)}{r_0}
\right)
\end{eqnarray}
is the Kullback-Leibler entropy of the ISI distribution with the
{\it frozen} rate $r(t)$ with respect to the unperturbed ISI
distribution with the rate $r_0$. The meaning of the result in Eq.
(\ref{result}) is as follows:  the {\it rate of information
transduction = spiking rate $\times$ relative change of the ISI
entropy}. Notably it does not depend of the time bin $\Delta \tau$.
The total information about signal is the time-integral of Eq.
(\ref{result}). Fig. \ref{Fig3} depicts the relative entropy for the
exponential distribution as a function of $r(t)/r_0$. Notice that
the information transferred per one spike can in principle exceed
one bit for a strong rate-modulation, in accordance with
\cite{Spikes}. This is because the information is transferred not
only via the  spike occurrence, but also with its {\it timing} (and
time is a continuous variable). For the weak signals ($qV_s<D$), the
information transfer per one spike, however, does not exceed one
bit, independently of the time resolution $\Delta\tau$.
%Here, our approach profoundly differs from the viewpoint accepted in \cite{Spikes}.

\subsection{Rate of information gain for a two-state process}

A generalization of the above result for the two-state Markovian
rate-modulated process $x(t)$ can also be obtained \cite{GH00}. The
Markovian two state dynamics is governed by the master equation
\begin{eqnarray}
\dot{p}_{1}(t)=
r_2(t)p_2(t)-r_1(t)p_1(t),\;\;p_2(t)=1-p_1(t)\;,
\end{eqnarray}
with the time-dependent transition rates $r_{1,2}(t)$ detailed e.g.
for SR gating in ion channels in Ref. \cite{GH00}. Using the present
notation the rate of information gain then emerges as
\begin{eqnarray}\label{two-state}
 \frac{d}{dt}K_{[t_0,t]}[x(t)|x_0(t)]=\sum_{i=1,2} r_i(t)
  K_{t}(\psi_{i}|\psi_{i}^{(0)})p_{i}(t)\;,
 \end{eqnarray}
where the relative entropy is given again by Eq. (\ref{rel}) for the
residence time distributions in the two states, reading
$\psi_{1,2}(\tau)=r_{1,2}(t)\exp[-r_{1,2}(t)\tau]$ with the frozen
rates $r_{1,2}(t)$, and $\psi_{1,2}^{(0)}(\tau)=
r_{1,2}^{(0)}\exp(-r_{1,2}^{(0)}\tau)$ are the rates in the absence
of the signal. In  \cite{GH00} this result has been applied to
investigate the problem of stochastic resonance in biological ion
channels \cite{Bezrukov} from an information theory perspective.
%One may conjecture that a result with a similar structure as one in Eq. (\ref{two-state})
%will hold also
%for arbitrary multi-state Markovian processes.

\section{Nonstationary SR}

\begin{figure}
\begin{center}
\resizebox{0.6\columnwidth}{!}{%
  \includegraphics{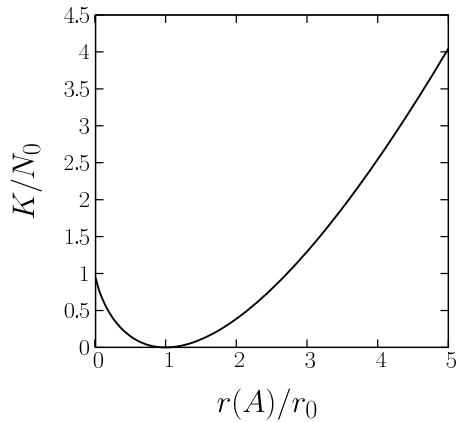}
}
\end{center}
\vspace{1cm}
%\vspace{5cm}       % Give the correct figure height in cm
\caption{Information gain $K$ per one spike of the undriven
Poisson process (in nats) as
function of the relative change of the spiking rate $r(A)/r_0$.} 
\label{Fig4}
\end{figure}

Now we are sufficiently  equipped in order to address the questions
posed in the abstract. We adhere  here to the Poisson model and
apply a transient step-like signal of the amplitude $A$ and the
duration $\tau_0$ depicted  with Fig. \ref{Fig1}, i.e. $V_s(t)=A$
for $t_0<t_{in}<t<t_{in}+\tau_0$,  $V_s(t)=0$  otherwise, and where
$t_{in}$ is the time instant when the signal is applied. The signal
can be either positive, $A>0$ (activating signal), or negative,
$A<0$ (inhibiting signal). In terms of the averaged number of
background spikes, $N_0=r_0\tau_0$, occurring within a typical time
interval of the duration $\tau_0$ in the absence of signal, the
total information gain is evaluated to read
\begin{eqnarray}
K=K_{[t_0,\infty]}[\xi(t)|\xi_0(t)]=
N_0 [r(A)/r_0]K(\psi|\psi_0),
\end{eqnarray}
where $r(A)=r(t)=const$ for $t_{in}<t<t_{in}+\tau_0$ and
$K(\psi|\psi_0)$ is given by Eq. (\ref{rel}) with $r(t)=r(A)$. This
is result after integrating Eq. (\ref{result}). Its structure is
illuminating; in terms of $N_0$ the signaling information transfer
involves both  the change of the  spiking rate with applied signal
$r(A)$ and  the relative change of the ISI-entropy. Interestingly,
for $r(A)/r_0<1$ (inhibiting signal) the information transferred per
one background spike cannot exceed one nat, cf. Fig. \ref{Fig4}. Put
differently, then each disappeared background spike bears no more
than one nat of information, in accord with intuition.

\begin{figure}
\begin{center}
\resizebox{0.7\columnwidth}{!}{%
  \includegraphics{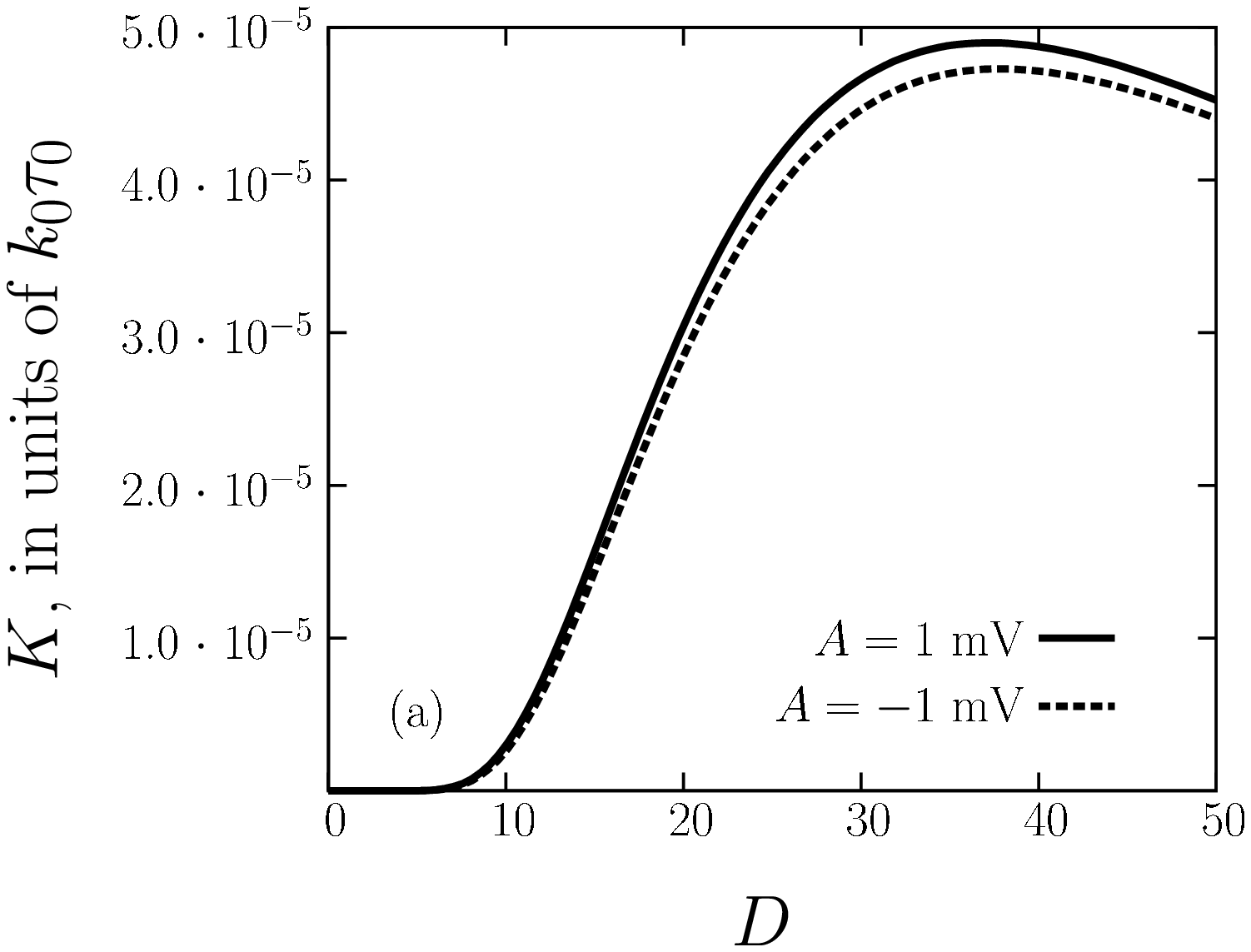}
}
\end{center}
\begin{center}
\resizebox{0.7\columnwidth}{!}{%
  \includegraphics{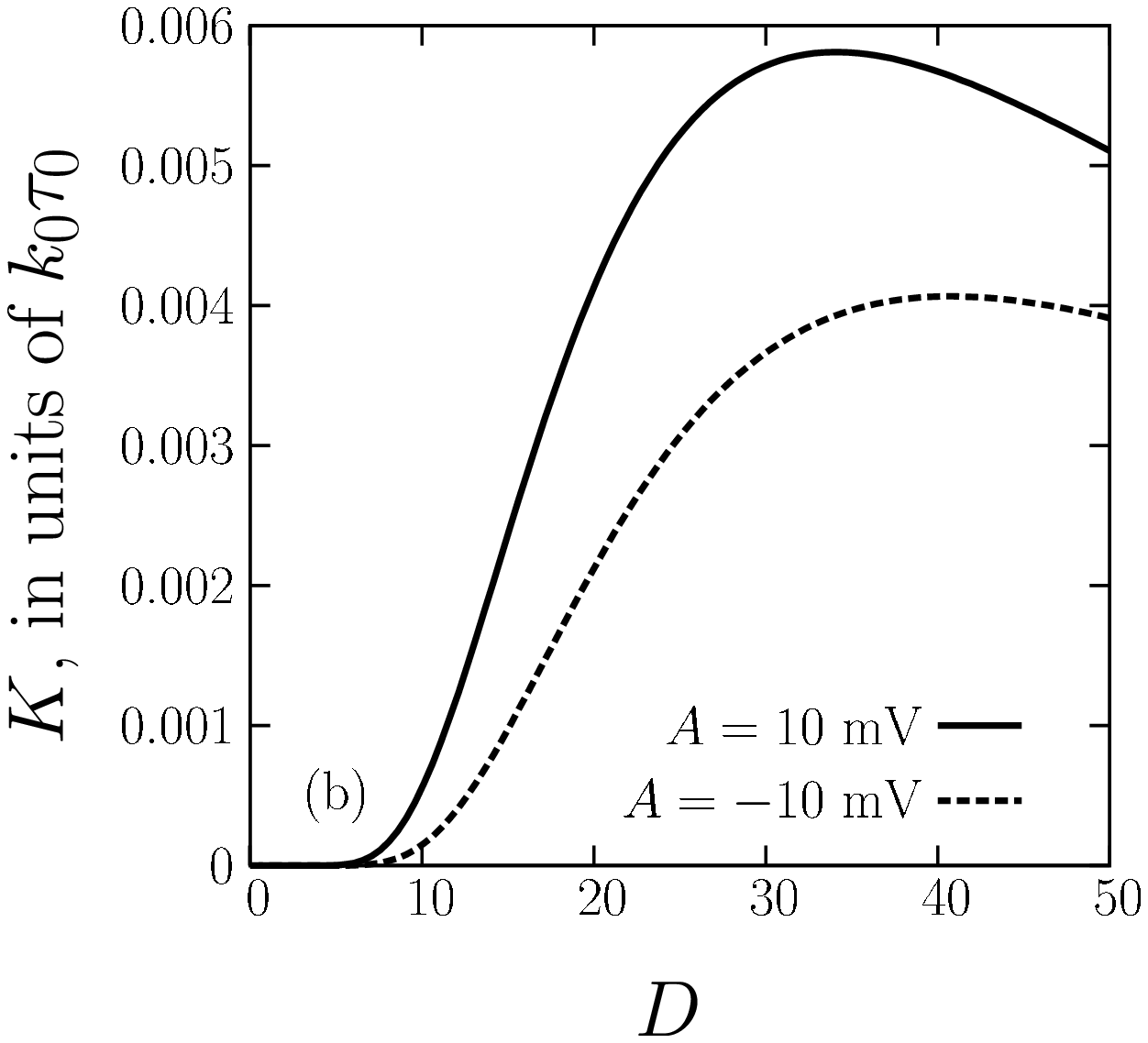}
}
\end{center}
\begin{center}
\resizebox{0.7\columnwidth}{!}{%
  \includegraphics{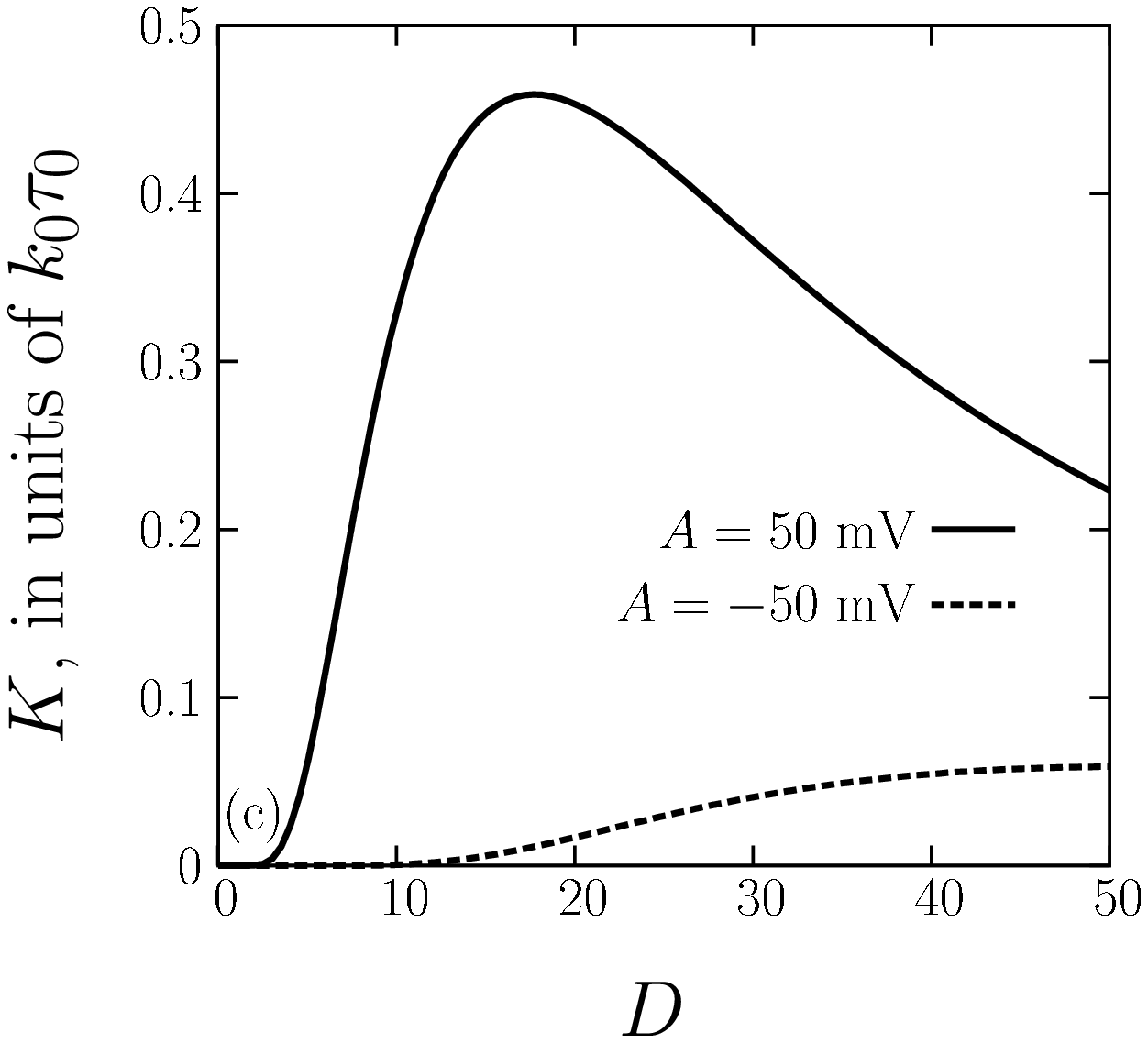}
}
\end{center}
\begin{center}
\resizebox{0.7\columnwidth}{!}{%
  \includegraphics{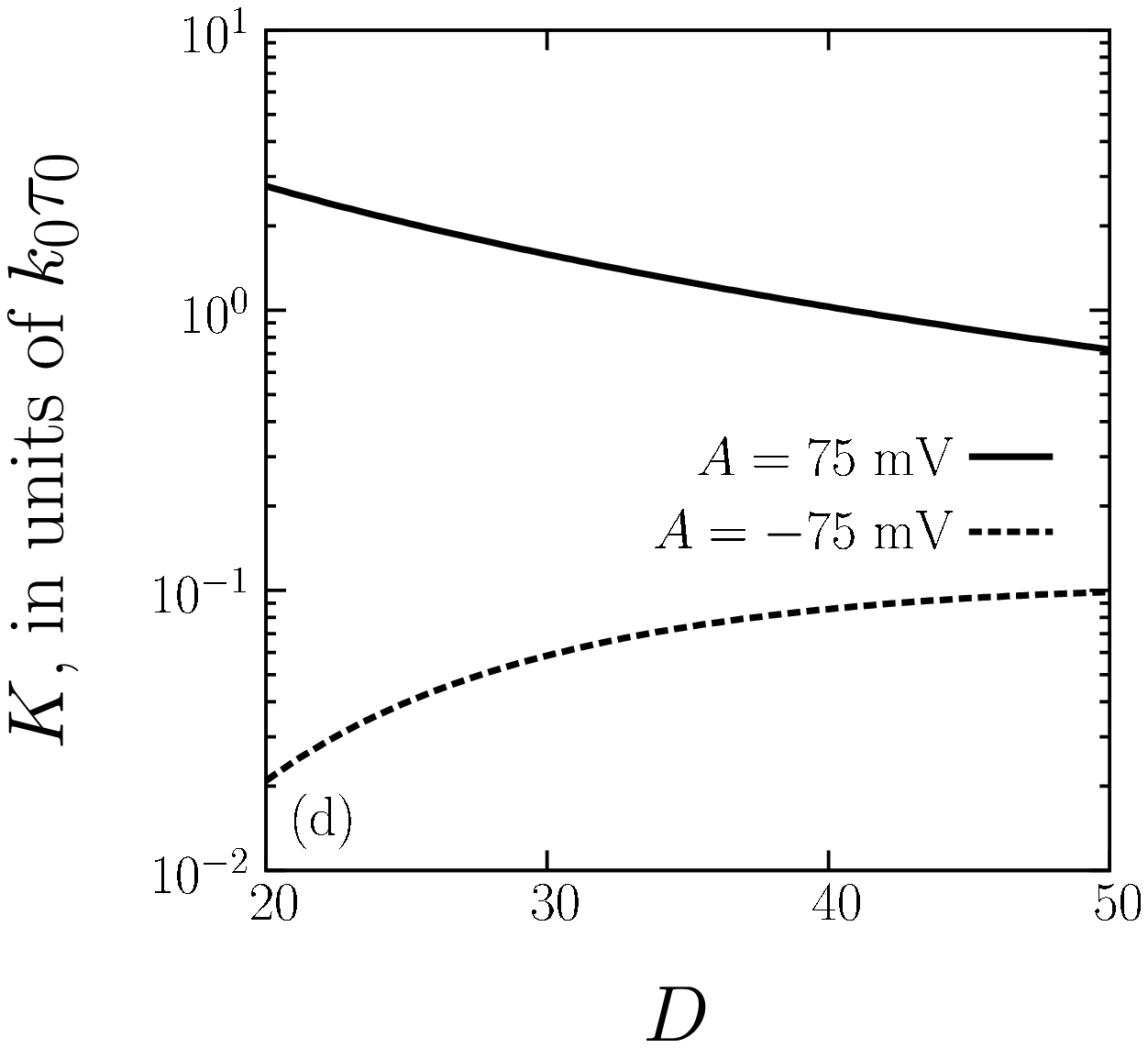}
}
\end{center}
%\vspace*{5cm}       % Give the correct figure height in cm
\caption{Information gain $K$ (in nats) versus the noise 
strength $D$ (in meV) 
in one of the basic SR models \cite{Review98,Model}, cf.
Eqs. (\ref{model1}), (\ref{model2}),  
for different values of the signal
strength $A$. Information gain is scaled in the  
units of $k_0\tau_0$. }
\label{Fig5}       % Give a unique label
\end{figure}

Furthermore, for the popular model in Eqs. (\ref{model1}),
(\ref{model2}), we obtain a practical  result, reading
\begin{eqnarray}\label{SR}
 K=k_0\tau_0 \exp\left(-\frac{U_0}{D}\right)
\left [1- \exp\left(\frac{qA}{D}\right)\left
(1- \frac{qA}{D}\right) \right ].
 \end{eqnarray}
For very weak signals, $q|A|\ll D$, it displays the well-known,
bell-shaped SR dependence on the noise intensity $D$ \cite{Review98}
\begin{eqnarray}
K \approx k_0\tau_0 \frac{(qA)^2}{D^2}
\exp\left(-\frac{U_0}{D}\right)\;,
\end{eqnarray}
being proportional to the SNR in the case of sinusoidal signal with
the same amplitude \cite{Review98,Goychuk01,Model}. The latter
result confirms the fact that the use of  SNR indeed describes
information transfer for weak signals. In order to detect a signal,
the total information gain should intuitively be no less than one
bit. This clearly poses a bound on the signal duration $\tau_0$
depending on its amplitude strength $A$. This bound can be found
from Eq. (\ref{SR}): Weak signals should last for a sufficiently
long time $\tau_0$ (many affected spikes accumulate the
corresponding information), otherwise such signals cannot be
detected within  environmental noise. In Fig. \ref{Fig5}, we depict
the information gain $K$ versus the noise strength $D$ for several
values of the signal strength $A$. The threshold barrier $U_0$ is
set to $U_0=75$ meV. Assuming that $q$ is equal to the elementary
charge, the signal strength is measured in mV.

The presence of wide sense  SR is clearly detectable in Fig.
\ref{Fig5}(a-c), for both activating and inhibiting signals. For an
activating signal of the threshold strength, $qA=U_0$, in Fig.
\ref{Fig5}(d), nonstationary SR disappears. The increase of the
information transfer by increasing the randomness of the background
process has an instructive  explanation. Namely, the increase of the
spontaneous spiking rate leads to more spikes occurring within the
signal duration $\tau_0$. They altogether transfer more information
about the signal. From the value of $K$ in Fig. \ref{Fig5}(a) one
can realize that many spikes are required in order to transfer
information of $K=1$ nat (approximately $0.7$ bit) about the
corresponding weak signal. Namely, one needs $k_0\tau_0>2\cdot
10^4$. This clearly poses a bound $\tau_0>2\cdot 10^4/k_0$ on its
duration. Assuming that $k_0=20\;{\rm ms}^{-1}$ (such that
$r_0\approx {\rm 1\; ms}^{-1}$ at $D=25$ meV), this yields the bound
$\tau_0>1$ sec, i.e. such a faint signal should last at least for
about 1000  spikes to become detectable. However, if to increase the
strength of the signal to 10 mV, the corresponding bound for
$\tau_0$ drops by  two orders of magnitude, as it can be deduced
from Fig. \ref{Fig5}(b). In other words, such a stronger input
signal (being, however, still much below the threshold) can be
detected already with a few spikes, in principle. Moreover,
nonstationary SR can help to detect such signals which would
otherwise far too short lived at a non-optimal noise intensity $D$.

\section{Conclusion}

In this work, we considered a  basic model for nonstationary SR,
i.e. for the case of deterministic but aperiodic signals of finite
duration. Information theory helped us to shed light on the very
possibility and the origin of such nonstationary SR, as well as
other critical issues such as the existence of a bound on the signal
duration $\tau_0$ versus its strength $A$. The obtained results
may be of a broad importance in the context of information
transduction in biological systems on the cellular level and in 
sensory systems.

\section{Acknowledgements}
This work has been supported by the Deutsche Forschungsgemeinschaft through
the SFB 486, and by the  German Excellence Initiative via the
Nanosystems Initiative Munich (NIM).

%
%
% BibTeX users please use
% \bibliographystyle{}
% \bibliography{}

\end{document}